\documentclass[12pt]{article}

\usepackage{amsmath, amssymb}
\usepackage{amsfonts}

\begin{document}


\def\a{\alpha}
\def\b{\beta}
\def\c{\varepsilon}
\def\d{\delta}
\def\e{\epsilon}
\def\f{\phi}
\def\g{\gamma}
\def\h{\theta}
\def\k{\kappa}
\def\l{\lambda}
\def\m{\mu}
\def\n{\nu}
\def\p{\psi}
\def\q{\partial}
\def\r{\rho}
\def\s{\sigma}
\def\t{\tau}
\def\u{\upsilon}
\def\v{\varphi}
\def\w{\omega}
\def\x{\xi}
\def\y{\eta}
\def\z{\zeta}
\def\D{{\mit \Delta}}
\def\G{\Gamma}
\def\H{\Theta}
\def\L{\Lambda}
\def\F{\Phi}
\def\P{\Psi}
\def\O{\Omega}
\def\S{\Sigma}
\def\V{\varPsi}
\newcommand{\EV}{ \,{\rm eV} }
\newcommand{\KEV}{ \,{\rm keV} }
\newcommand{\MEV}{ \,{\rm MeV} }
\newcommand{\GEV}{ \,{\rm GeV} }
\newcommand{\TEV}{ \,{\rm TeV} }

\def\o{\over}
\newcommand{\sla}[1]{#1 \llap{\, /}}
\newcommand{\beq}{\begin{eqnarray}}
\newcommand{\eeq}{\end{eqnarray}}
\newcommand{\bea}{\begin{eqnarray}}
\newcommand{\eea}{\end{eqnarray}}
\newcommand{\gsim}{ \mathop{}_{\textstyle \sim}^{\textstyle >} }
\newcommand{\lsim}{ \mathop{}_{\textstyle \sim}^{\textstyle <} }
\newcommand{\vev}[1]{ \left\langle {#1} \right\rangle }
\newcommand{\bra}[1]{ \langle {#1} | }
\newcommand{\ket}[1]{ | {#1} \rangle }
\newcommand{\1}{\mbox{1}\hspace{-0.25em}\mbox{l}}
\newcommand{\cG}{{\cal G}}
\newcommand{\cN}{{\cal N}}
\newcommand{\cB}{{\cal B}}
\newcommand{\Mpl}{M_{\mbox{\scriptsize pl}}}


\baselineskip 0.7cm

\begin{titlepage}

\begin{flushright}
YITP-10-73 \\
IPMU 10-0143
\end{flushright}

\vskip 1.35cm
\begin{center}
{\large \bf
Gravitational Supersymmetry Breaking
}
\vskip 1.2cm
Izawa K.-I.$^{1,2}$, T.~Kugo$^1$, and T.T.~Yanagida$^2$
\vskip 0.4cm

{\it $^1$Yukawa Institute for Theoretical Physics, Kyoto University,\\
     Kyoto 606-8502, Japan}

{\it $^2$Institute for the Physics and Mathematics of the Universe, University of Tokyo,\\
     Chiba 277-8568, Japan}

\vskip 1.5cm

\abstract{
We consider supersymmetry breaking models
with a purely constant superpotential in supergravity.
The supersymmetry breaking is induced for the vanishing
cosmological constant. 
As a hidden mediation sector of supersymmetry breaking,
it naturally leads to a split spectrum
in supersymmetric standard model.
We also point out possible utility of our setup
to construct nonlinear sigma model and/or Fayet-Iliopoulos-like
term in broken supergravity.}

\end{center}
\end{titlepage}

\setcounter{page}{2}


It naively seems that perturbative gravity as an effective theory
provides only small corrections
to low-energy particle physics on the flat background and thus it would not
essentially alter the physical contents such as phase structure
from those of the corresponding particle physics without gravity.
However, as for supergravity in superspace, such a naive guess
is not necessarily true. Simple-minded decoupling of the supergravity multiplet
sometimes results in qualitatively different physics on the flat background. 

In this paper, we consider supersymmetry (SUSY) breaking models
with a purely constant superpotential in supergravity.
Although the constant superpotential has no physical meaning
in the case of rigid SUSY without gravity,
the SUSY breaking is induced for the vanishing
cosmological constant in the case of supergravity. 
As a hidden mediation sector of SUSY breaking,
it naturally leads to a split spectrum
in supersymmetric standard model.
We also point out possible utility of our setup
to construct nonlinear sigma model and/or Fayet-Iliopoulos-like
$D$-term in broken supergravity.

Let us adopt a purely constant superpotential%
\footnote{
Classically the K\"ahler-Weyl transformation can achieve the constant
superpotential without loss of generality,
while quantum-mechanically it is generically anomalous.}
\beq
 W=c \neq 0,
\eeq
and a K\"ahler potential $K$ of chiral superfields $\F_i$,
where the subscript $i$ denotes a flavor index.
Our crucial premise is that the K\"ahler potential $K$
is tuned to make the cosmological
constant vanishing so that SUSY is spontaneously broken
with its breaking scale
solely determined by the constant $W=c$ or the gravitino mass.%
\footnote{Concrete structure of the K\"ahler potential
was investigated by Hebecker
\cite{Heb}
in the case of a single chiral superfield with the constant
superpotential.}
We now proceed to investigate physical contents of this setup.

With the graviton field and all the fermionic fields vanishing,
the supergravity action takes the form
\beq
 \int \! d^4x\,d^2\h d^2\h^* \ \v \v^*\O
 +\left(\int \! d^4x\,d^2\h \ \v^3W+\mbox{h.c.}\right)+\cdots,
\eeq
where
$\O=-3\exp(-K/3)$ with the reduced Planck scale unity
and
the chiral compensator $\v=1+\h^2F$.
Here the compensator lowest component is gauge-fixed to be one
for simplicity.
The ellipses stand for extra derivative terms for (the lowest
components of) $\Phi_i$ which come from the vector auxiliary field
\cite{Kug}.

The algebraic equations of motion for the auxiliary fields
$F$ $(\equiv F_\v)$ and $F_{\F_i}$
determine them in terms of the dynamical fields $\F_i$,
which yield the (non-canonically normalized) scalar potential
\beq
 V=-3cF,
\eeq
as will be shown just below.
Hence the vanishing of the cosmological constant results in
the vanishing of the compensator auxiliary field $F$ in the vacuum.
In particular, it means an important implication that
the so-called anomaly mediation
\cite{Ran}
does {\em not} occur.

More generally, we can show that the vanishing $F$ is realized in the case
of vanishing $F_W$ for a generic superpotential $W$.
The action
\beq
 \int \! d^4x\,d^2\h d^2\h^* \ \v \v^*\O
 +\left(\int \! d^4x\,d^2\h \ \v^3W+\mbox{h.c.}\right),
\eeq
with the chiral compensator
$\v=1+\h^2F$,
yields the scalar potential
\beq
 &&-V=|F|^2\O+F\O^{\bar i}F_{\bar i}^*+F^*\O^iF_i+\O^{{\bar i}j}F_{\bar i}^*F_j
    +3FW+F_W+3F^*W^*+F_W^* \nonumber \\
 &&\qquad = F^*(F\O+\O^iF_i+3W^*)+F_{\bar i}^*(F\O^{\bar i}+\O^{{\bar i}j}F_j+W^{*{\bar i}})
    +3FW+F_W,
\eeq
with $F_i=F_{\F_i}$ and $F_W=W^iF_i$,
where the superscripts $i$ to $\Omega$ and $W$ denote partial derivatives
with respect to $\Phi_i$ and  the repeated indices are summed over.

The algebraic equations of motion for the auxiliary fields $F^*$
and $F_{\bar i}^*$ read, respectively,
\beq
 & &F \O + \O^i F_i + 3W^* = 0,
 \label{eom} \\
 & &F \O^{\bar i} + \O^{\bar i j}F_j + W^{* \bar i} = 0,
\eeq
which lead to $V= -3FW - F_W.$

With this, we see that the demand $\vev{V}=\vev{F_W}=0$ generally implies
$\vev{F}=0$ under $\vev{W} \neq 0$.
That is, the vacuum value of the compensator auxiliary field vanishes.
When $W$ is a constant ($W=c$), in particular, then $F_W=0$ since
$W^i=0$, which results in $V=-3cF$ as advocated above.
Eq.(\ref{eom}) also implies that,
when $\vev{F}=0$, SUSY is spontaneously broken since
$\vev{W} \neq 0$ requires $\vev{\O^iF_i} \neq 0$  so that $\vev{F_i} \neq 0$
for some $i$.

Let us turn to phenomenological implications of 
our setup with the constant superpotential ($W=c$).
Since we assume the unique input scale $c$ as the origin of SUSY breaking,
we can make order estimation of soft masses
based on possible forms of effective operators.
Under the vanishing compensator auxiliary field ($F=0$),
we have three possible contributions to the gaugino mass term
or the lowest component of ${\cal W}^\a{\cal W}_\a$
with the reduced Planck scale as the cutoff.

\noindent
$i)$ {gravitational mediation}:

The contribution to the gaugino masses
is estimated to be of order $c^3$ due to effective operators
\beq
 \int \! d^2\h d^2\h^* \ |\F_i|^2W^*{\cal W}^\a{\cal W}_\a,
\eeq
of the neutral R charge
\cite{Ark}.
In other words, the maximum $\vev{F_{\Phi_i}}$ is naturally of order the
gravitino mass $m_{3/2}$, so that the gaugino masses are expected to be
given by ${\cal O}(m_{3/2}^3)$.
To obtain the gaugino masses at the order of 100GeV, we
need to choose $m_{3/2} \simeq 10^{13}$GeV.

\noindent
$ii)$ {moduli mediation}:

The contribution to the gaugino masses
is estimated to be of order $c$ or less due to effective operators
\beq
 \int \! d^2\h \ \F_i{\cal W}^\a{\cal W}_\a.
\eeq

The above two contributions constitute gravity mediation of SUSY
breaking suppressed by the Planck scale, while gauge mediation
\cite{Giu}
can incorporate a lower scale of SUSY breaking mediation as follows.

\noindent
$iii)$ {gauge mediation}:

The contribution to the gaugino masses
is estimated to be of order $c/M$ or less due to effective operators
\beq
 \int \! d^2\h \ {\F_i \o M} {\cal W}^\a{\cal W}_\a,
\eeq
where $M$ denotes the scale of gauge mediation.

In the three contributions, the first one is ubiquitous
in contrast to the remaining two.
Without the moduli and gauge mediation, the gravitational mediation
naturally leads to a split spectrum
in supersymmetric standard model due to the vanishing anomaly mediation,
where gaugino masses are many orders of magnitude lighter than scalar masses.

One peculiar realization may appear as a tiny bino mass%
\footnote{The other gauginos
are assumed to obtain larger masses from other contributions.}
less than 10eV,
which avoids cosmological problems due to thermal bino abundance.
Heavier bino mass would
necessitate tuning for co-annihilation with slepton
in order to achieve efficient bino
annihilation in the early universe.

Even in the presence of other dominant contributions to the bino
mass, the absence of the anomaly mediation may have some virtues.
For instance, sizable anomaly mediated contribution might induce
CP-violating phase difference among different origins of the gaugino masses
\cite{End},
which seems undesirable in view of experimentally
limited CP-violating effects. 

Finally a few comments are in order.
As a by-product, the present K\"ahler potential can be utilized to
construct nonlinear sigma model and/or Fayet-Iliopoulos-like
$D$-term in broken supergravity.

For simplicity, let us restrict ourselves to the single chiral superfield $\Phi$
with $\O(\f)$ and $W=c \neq 0$, where $\f=\Phi+\Phi^*$.
Then, in view of Eq(\ref{eom}), the vanishing cosmological constant requires
\beq
 \vev{\O'(\f) F_\Phi} + 3c^*=0,
\eeq
which implies $\vev{\O'(\f)} \neq 0$.

When we have the K\"ahler potential $\cG$ of a nonlinear sigma model
with $\vev{\cG}=0$ in rigid SUSY
\cite{Ban},
we can obtain a corresponding
nonlinear sigma model in broken supergravity
in terms of the above setup with $\O(\f+\cG)$,
since its expansion
\beq
 \O(\f+\cG)=\O(\f)+\O'(\f)\cG+{1 \o 2}\O''(\f)\cG^2+ \cdots
\eeq
contains the leading term $\O'(\f)\cG$ under $\vev{\O'(\f)} \neq 0$,
provided SUSY is almost intact in the sigma model sector. 

Even if a symmetry transformation in rigid SUSY varies $\cG$
by a holomorphic function plus its complex conjugate,
the variation can be absorbed in $\f$,
and the symmetry is maintained in the supergravity.

We may also realize a Fayet-Iliopoulos-like $D$-term in supergravity
\cite{Kom}
by replacing the above $\cG$ with a $U(1)$ real superfield,%
\footnote{If instead we replace the above $\cG$ with $\cG$ plus a $U(1)$ real
superfield, the chiral superfield $\Phi$, which contains a massless scalar,
becomes a gauge degree of freedom so that the physical massless modes
are exclusively contained in those of the original sigma model.}
whose gauge transformation has a similar property of holomorphy.

\section*{Acknowledgements}

This work is supported by the Grant-in-Aid for Yukawa International
Program for Quark-Hadron Sciences, the Grant-in-Aid
for the Global COE Program "The Next Generation of Physics,
Spun from Universality and Emergence", and
World Premier International Research Center Initiative
(WPI Initiative), MEXT, Japan.
TK is also partially supported by a Grant-in-Aid for Scientific
Research (B) (No.\ 20340053) from the Japan Society for the Promotion
of Science.

\end{document}